\documentclass[11pt,twocolumn,showpacs,floatfix,superscriptaddress,nofootinbib,amssymb,amsmath,aps,prc]{revtex4-2}

\usepackage{CJK}
\usepackage[dvips]{graphicx}
\usepackage{epstopdf} 
\usepackage{latexsym,amssymb,amsmath,epsfig,bm,times,psfrag,subfig}
\usepackage{color}
\usepackage{mathptmx}
\usepackage{dcolumn}
\usepackage[bookmarksnumbered,bookmarksopen,colorlinks,citecolor=blue,linkcolor=blue]{hyperref}
\usepackage{placeins}   
\newcommand{\be}{\begin{equation}}
\newcommand{\ee}{\end{equation}}
\newcommand{\bear}{\begin{eqnarray}}
\newcommand{\eear}{\end{eqnarray}}
\newcommand{\bB}{{\bm B}}

\newcommand{\bb}{{\bm b}}
\renewcommand{\bm}{\boldsymbol}

\begin{document}
\begin{CJK*}{UTF8}{gbsn}

\title{Event-by-event analysis of chiral charge separation in $p^{\uparrow}+$Au collisions within an improved AMPT model}

\author{Chen Gao (高晨)}
\affiliation{School of Physics, Huazhong University of Science and Technology, Wuhan 430074, China}

\author{Gui-Zhen Wu (吴桂珍)}
\affiliation{School of Physics, Huazhong University of Science and Technology, Wuhan 430074, China}

\author{Yi Xu (许易)}
\affiliation{School of Physics, Huazhong University of Science and Technology, Wuhan 430074, China}

\author{Wei-Tian Deng (邓维天)}
\thanks{The corresponding author}
\email{dengwt@hust.edu.cn}
\affiliation{School of Physics, Huazhong University of Science and Technology, Wuhan 430074, China}

\begin{abstract}
We study the charge separation induced by chiral magnetic effect (CME) in $p^{\uparrow}$+Au collisions at $\sqrt{s_{NN}}=200$ GeV event-by-event with an improved string-melting AMPT model. In this model, an impact-parameter-dependent formation-time delay is introduced to reduce the peak Bjorken energy density from $\sim$2 to $\sim$0.3 GeV/fm$^3$ and reproduces the illiptic flow $v_2(p_T)$ measured by PHENIX in $p+$Au simultaneously. The event-by-event CME source, whose quark momentum-exchange fraction scales with $|\mathbf{B}|_{\mathrm{event}}/|\mathbf{B}|_{\mathrm{max}}$ and is capped at the 7\% value from Au+Au, transmits the field geometry directly to the observable. We find that the initial correlator shows a clear scheme hierarchy tracking $B^2$. The parton cascade preserves $\sim$80--90\% of the signal, while coalescence and ART hadronic rescattering dissipate the bulk, leaving $\sim$10--30\% in the final state. The $\gamma_{OS}/\gamma_{SS}$ splitting remains visible. These results support the use of the inter-scheme difference $\Delta\gamma_{\mathrm{IV}}-\Delta\gamma_{\mathrm{II}}$ as an essentially background-free CME observable as we proposed in previous work.
\end{abstract}
\maketitle
\end{CJK*}

\section{Introduction\label{sec:intro}}

In off-central relativistic heavy-ion collisions the QCD matter reaches temperatures well above the deconfinement transition, producing a quark-gluon plasma (QGP), and a transient electromagnetic field of unprecedented strength is generated by the spectator protons~\cite{Rafelski:1975rf,Skokov:2009qp,Bzdak:2011yy,Voronyuk:2011jd,Deng:2012pc,Deng:2014uja,Inghirami:2016iru,Yan:2021zjc}. Together with the topological fluctuations of the QCD vacuum, this strong field can drive a macroscopic separation of electric charge along the field direction, the so-called chiral magnetic effect (CME)~\cite{Kharzeev:1998kz,Kharzeev:2004ey,Kharzeev:2007jp,Fukushima:2008xe,Kharzeev:2015znc,Liu:2020ymh,Kharzeev:2009fn,Fukushima:2010vw,Basar:2010zd}, which is one of the few collider observables sensitive to local $\mathcal{P}$ and $\mathcal{CP}$ violation in QCD.

To probe the resulting charge separation, the three-point azimuthal correlator
\be
\gamma_{\alpha\beta}=\langle \cos (\phi_{\alpha}+\phi_{\beta}-2\Psi_{RP})\rangle
\label{Eq-gamma}
\ee
is commonly used~\cite{Voloshin:2004th}, where $\phi_{\alpha,\beta}$ are the azimuthal angles of two charged hadrons and $\Psi_{RP}$ is the reaction-plane angle. The charge-dependent difference
\be
\Delta\gamma=\gamma_{OS}-\gamma_{SS}
\label{Eq-Delta-gamma}
\ee
cancels charge-independent backgrounds~\cite{Kharzeev:2015znc}, and the CME contribution to it scales as~\cite{Bloczynski:2013mca}
\be
\Delta\gamma_{\mathrm{CME}} \propto B^{2}\cos\!\big[2(\Phi_{B}-\Psi_{RP})\big].
\label{Eq-Delta-gamma-CME}
\ee
However, $\Delta\gamma$ also picks up a sizable, $v_2$-correlated background of conventional QCD origin~\cite{Xu:2017zcn,STAR:2013zgu,Wang:2016iov,Ajitanand:2010rc}, and the recent isobar measurement at RHIC has shown that this background is so large in heavy-ion collisions that an unambiguous identification of the CME is still out of reach~\cite{Deng:2016knn,STAR:2021mii}. In small systems, the CMS measurement of charge-dependent azimuthal correlations in $p$+Pb at the LHC has further indicated that the observed signal can be largely accounted for by $v_2$-related background~\cite{CMS:2016wfo}, reinforcing the difficulty of isolating the CME with the standard $\gamma$ correlator alone.

In a series of previous works we have argued that this difficulty can be circumvented by using a small system in which one beam is replaced by a transversely polarized proton, $p^{\uparrow}+A$~\cite{Zhang:2021jrc,Wu:2024vcd,Xu:2025cme}. Although smaller and shorter-lived, such systems are by now known to develop a partonic phase: the PHENIX geometry scan of $p+$Au, $d+$Au, and $^{3}$He$+$Au at $\sqrt{s_{NN}}=200$~GeV has shown that the elliptic and triangular flow coefficients faithfully track the initial geometric eccentricities~\cite{PHENIX:2018lia}, and AMPT-based transport calculations reproduce both the magnitude and the ordering of these flows~\cite{OrjuelaKoop:2015jss}. The asymmetric charge distribution inside the polarized proton makes the electromagnetic field at the overlap region depend strongly on the orientation of the impact parameter relative to the polarization axis, so that four collision-geometry schemes (defined precisely in Sec.~\ref{sec:method}) probe four well-separated values of $B^2$. We have shown that $\Delta\gamma_{\mathrm{CME}}$ differs substantially between these schemes~\cite{Wu:2024vcd}, that comparing schemes lets us cancel the flow-driven background and isolate the CME signal, and that the background contamination of $\Delta\gamma$ in $p+$Au is in fact negligibly small~\cite{Xu:2025cme}. In Ref.~\cite{Xu:2025cme} we further tracked how an initially injected charge separation survives parton cascade and hadronic rescattering, finding that of order $30$--$50\%$ of the signal survives to the final state when a fixed $7\%$ exchange is applied.

In this paper we extend this programme in two directions. First, instead of attaching a fixed momentum-exchange fraction to each event as in Ref.~\cite{Xu:2025cme}, we implement an event-by-event CME source in which the exchange fraction scales linearly with the local magnetic-field strength $|\bB|$ at the overlap centre. The overall normalisation is fixed by requiring that the most magnetized events reach the 7\% value that reproduces the $\gamma_{SS}$ correlator measured in 30--50\% Au+Au collisions at $\sqrt{s_{NN}}=200$~GeV, while the four geometric schemes pick out different distributions of $|\bB|$ and hence different average CME strengths. Second, the standard AMPT setup applied to $p+$Au at these energies produces a Bjorken energy density that exceeds the values expected for a small system by almost an order of magnitude. Following Zhao~\emph{et al.}~\cite{Zhao:2024lpc}, who introduced an impact-parameter-dependent hadron formation-time delay to address the same problem in $^{16}$O$+^{16}$O collisions, we adopt and re-tune an analogous $\Delta\tau(b)$ for $p+$Au. Combined with the partonic improvements of Ref.~\cite{Zhang:2021ggt}, which underpins our AMPT version, the modified setup brings the peak energy density into the $\sim 0.3$~GeV/fm$^3$ range and simultaneously reproduces the elliptic flow $v_2(p_T)$ measured by PHENIX.

The paper is organized as follows. Section~\ref{sec:method} describes the improved AMPT framework, including the formation-time delay and the event-by-event CME implementation. Section~\ref{sec:validation} validates the model against the Bjorken energy density and the PHENIX $v_2(p_T)$ data. Section~\ref{sec:results} presents the charge-separation correlator as a function of $P_+$ and $\Delta\eta$ for the four schemes. We summarize in Sec.~\ref{sec:summary}.

\FloatBarrier
\section{Model and Method\label{sec:method}}

\subsection{The AMPT model and collision geometry}

We use the string-melting version of the AMPT model~\cite{Zhang:1999bd,Lin:2004en}, with the local-nuclear-scaling refinement of the Lund string-fragmentation parameter $b_L$ and the mini-jet cutoff $p_0$ introduced in Ref.~\cite{Zhang:2021ggt}, to simulate $p+$Au collisions at $\sqrt{s_{NN}}=200$~GeV. An event proceeds through four stages: (i) initial conditions generated by HIJING~\cite{Wang:1991hta,Gyulassy:1994ew}, with excited strings melted into partons; (ii) parton transport by the Zhang Parton Cascade (ZPC)~\cite{Zhang:1997ej}; (iii) hadronization through quark coalescence~\cite{Lin:2002gc,Lin:2001zk}; and (iv) hadronic rescattering described by A Relativistic Transport (ART) model~\cite{Li:1995pra}.

The proton moves along $+z$ and the Au nucleus along $-z$, each with $E=100$~GeV per nucleon. The proton is transversely polarized along $+x$. Following Ref.~\cite{Wu:2024vcd}, four collision-geometry schemes are defined by the orientation of the impact parameter $\bb$ relative to the polarization axis. In each event the azimuthal angle $\Psi_{RP}$ is also defined by the oritation of $bb$, which is sampled randomly within a $\pi/2$-wide quadrant centred on the typical direction of the scheme:
\begin{itemize}
\item Scheme I: $\Psi_{RP}\in(-\pi/4,\,\pi/4)$ (typical oritation of $\bb$ being along $+x$),	
\item Scheme II: $\Psi_{RP}\in(\pi/4,\,3\pi/4)$ (typical oritation of $\bb$ being  along $+y$),
\item Scheme III: $\Psi_{RP}\in(3\pi/4,\,5\pi/4)$ (typical oritation of $\bb$ being  along $-x$),
\item Scheme IV: $\Psi_{RP}\in(-3\pi/4,\,-\pi/4)$ (typical oritation of $\bb$ being  along $-y$).
\end{itemize}
Because the four quadrants together tile the full $2\pi$ exactly once, each event is assigned unambiguously to a single scheme. 
\begin{figure}[!htbp]
	\centering
	\includegraphics[width=\linewidth,clip]{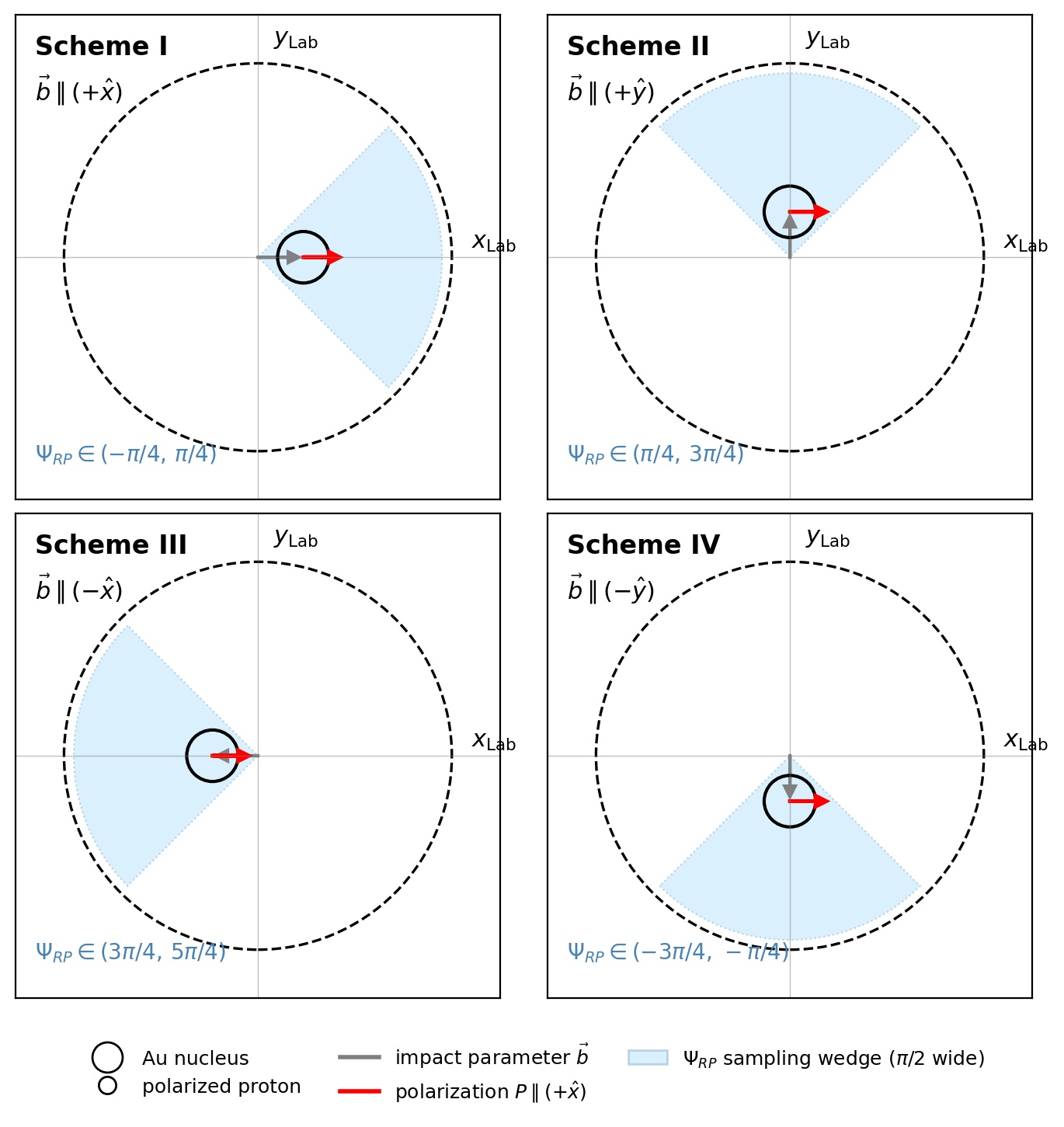}
	\caption{Illustration of the four collision-geometry schemes in the laboratory $x_{\rm Lab}$--$y_{\rm Lab}$ plane. The dashed circle represents the Au nucleus, the small solid circle the polarized proton, the red arrow its polarization direction $P\parallel +\hat{x}$, and the gray arrow the impact parameter $\bb$ pointing from the Au centre to the proton centre. The light-blue wedge marks the $\pi/2$-wide range over which $\Psi_{RP}$ is uniformly sampled within each scheme.}
	\label{fig:4scheme-geometry}
\end{figure}

In this work, any results within each scheme are averaged over events with random fluctuations of the reaction plane as a real experiment. The magnitude of impact parameter $|\bb|=b=1.5$~fm is fixed, in order to concentrate on the gamma-correlator analysis, corresponding approximately to the 0--5\% most central bin in $p+$Au.

\subsection{Hadron formation-time delay\label{sec:ftime}}

In our previous work Ref.~\cite{Xu:2025cme}, the parton cross section of $\sigma=10$~mb is used in the stage II of standand AMPT model. This value is appropriate for heavy ion collision system but drives the energy density of small and median collision system to unphysical large magnetude~\cite{Zhao:2024lpc}. In the present work we use $\sigma=0.3$ and $0.7$~mb, more in line with the partonic mean free path expected in $p+A$  small system. 
In the ZPC module, elastic $gg\to gg$ scattering is described by a Debye-screened $t$-channel gluon exchange with differential cross section $d\sigma/dt=9\pi\alpha_s^{2}/[2(t-\mu^{2})^{2}]$, which in the high-energy limit $s\gg\mu^{2}$ integrates to~\cite{Zhang:1997ej,Lin:2004en}
\be
\sigma_{gg}\;=\;\frac{9\pi\alpha_{s}^{2}}{2\mu^{2}},
\label{Eq-sigma-zpc}
\ee
where $\alpha_{s}$ is the strong coupling and $\mu$ is the Debye screening mass. In the standard AMPT parametrisation $\alpha_{s}$ is fixed and the total cross section can be tuned by adjusting $\mu$, so that $\mu^{2}\propto 1/\sigma$. Throughout this work we take $\alpha_{s}=0.33$; the two cross sections then correspond to $\mu=7.612$~fm$^{-1}$ ($\sigma=0.3$~mb) and $\mu=4.689$~fm$^{-1}$ ($\sigma=0.7$~mb). 

Even with these reduced cross sections, the standard AMPT setup still over-produces the Bjorken energy density (see Sec.~\ref{sec:validation}), because too many of the initial partons hadronize early and concentrate their transverse energy in a small longitudinal slice at midrapidity. Following the procedure introduced for $^{16}$O$+^{16}$O collisions in Ref.~\cite{Zhao:2024lpc}, we therefore add an impact-parameter-dependent delay $\Delta\tau(b)$ to the hadron formation time $\tau_{f}$ assigned in the quark-coalescence step. The delay enters as $\tau_{f}\to\tau_{f}+\Delta\tau(b)$ and is applied to every hadron produced from the partonic system; the parton-cascade dynamics in ZPC are left untouched. The net effect is to push the rapid pile-up of transverse energy at $y=0$ later in proper time, so that $\varepsilon_{Bj}(\tau)$ peaks at a lower value, without changing the integrated soft-particle yields.

We parametrize the delay with a sigmoid form,
\be
\Delta\tau(b) = \frac{A}{1 + B\, e^{b/C}},
\label{Eq-dtau}
\ee
where
\begin{itemize}
\item $A$ (in fm/$c$) is the overall amplitude of the delay; since $B\ll 1$ in our parametrisation, the small-$b$ plateau $\Delta\tau(0)=A/(1+B)$ is essentially equal to $A$, which therefore sets the maximum delay applied at central impact parameters,
\item $C$ (in fm) is the characteristic impact-parameter scale over which $\Delta\tau(b)$ falls off; the larger $C$, the more slowly the delay decreases with increasing $b$,
\item $B$ (dimensionless) controls the position of the inflection point through $\Delta\tau(b_*)=A/2$ at $b_*=-C\ln B$; smaller $B$ pushes the fall-off region to larger $b$ and thus widens the central plateau.
\end{itemize}
The three parameters are tuned by hand so that the peak Bjorken energy density at central rapidity stays below $\sim 0.3$~GeV/fm$^3$ in the most central bin while the $b$-dependence of the energy-density fall-off is reproduced; for the two parton cross sections used here we find
\begin{itemize}
\item $\sigma=0.3$~mb: $A=1.20$~fm/$c$, $B=0.0143$, $C=0.60$~fm,
\item $\sigma=0.7$~mb: $A=1.05$~fm/$c$, $B=0.0143$, $C=0.50$~fm.
\end{itemize}

\subsection{Event-by-event CME implementation\label{sec:cme}}

In Ref.~\cite{Xu:2025cme} a fixed 7\% momentum exchange was applied uniformly to each event regardless of the local magnetic field. The local $B$ field at the overlap centre, however, fluctuates strongly from event to event and depends on the collision-geometry scheme, so a fixed exchange fraction is at best approximation of the underlying CME dynamics.

We compute the electromagnetic field at the overlap centre event by event from the Lienard--Wiechert potential, using the three-dimensional asymmetric charge profile for the polarized proton and symmetric profiles for the nucleons in the Au nucleus, as in Refs.~\cite{Zhang:2021jrc,Wu:2024vcd}. The CME-induced charge separation is then injected at the parton level by interchanging the $p_y$ components of pairs of co-moving quark and antiquark whose momentum components along $\bB$ have opposite signs, mimicking the $\bB$-directed chirality-dependent momentum imbalance of the CME: the $p_y$ of an upward-moving $\bar{u}$ is swapped with that of a downward-moving $u$, and similarly for the $d/\bar{d}$ pair~\cite{Ma:2011uma}.

The fraction $f$ of such quark pairs that is exchanged in a given event is taken to scale linearly with the magnitude of the magnetic field at the overlap centre,
\be
f \;=\; 7\% \times \frac{|\bB|_{\mathrm{event}}}{|\bB|_{\mathrm{max}}},
\label{Eq-fraction}
\ee
where $|\bB|_{\mathrm{event}}$ is the magnetic-field strength at the overlap centre in the current event and $|\bB|_{\mathrm{max}}\simeq 5.6\,m_{\pi}^{2}/e$ is the largest value found across the full sample, taken over all four schemes. The 7\% ceiling is fixed by reproducing the $\gamma_{SS}$ correlator in 30--50\% Au+Au collisions at $\sqrt{s_{NN}}=200$~GeV~\cite{Xu:2025cme,STAR:2009wot,STAR:2009tro}. Because the four schemes pick out very different distributions of $|\bB|$, Eq.~(\ref{Eq-fraction}) automatically generates four distinct average exchange fractions $\langle f\rangle_{\mathrm{I,II,III,IV}}$, with scheme IV the largest and scheme II the smallest, in line with the geometric prediction $\Delta\gamma_{\mathrm{CME}}\propto B^{2}\cos[2(\Phi_{B}-\Psi_{RP})]$.

\subsection{Implementation in the AMPT code}

The three ingredients introduced above are implemented as local modifications of the public AMPT-SM source code, called as improved AMPT model in this work. The four collision-geometry schemes are realised in \texttt{hijing1.383\_ampt.f}: in each event the reaction-plane angle is drawn as $\Psi_{RP}=(\pi/2)\,\xi+\Psi_{0}$, with $\xi$ a uniform random number on $(-1/2,1/2)$ and $\Psi_{0}=0,\pi/2,\pi,3\pi/2$ selecting schemes I--IV. The CME source is implemented inside ZPC (\texttt{zpc.f}): the change subroutine first calls the precomputed Lienard--Wiechert table to obtain $|\bB|_{\mathrm{event}}$, evaluates the per-event fraction $f$ of Eq.~(\ref{Eq-fraction}) capped at unity, applies a Knuth shuffle to the lists of co-moving $u$, $\bar{u}$, $d$, $\bar{d}$ quarks, and then exchange the $p_y$ of the first $[f\,N_{q}]$ pairs (rounded to the nearest integer) in each flavour list. Finally, the hadron formation-time delay of Eq.~(\ref{Eq-dtau}) is added in \texttt{linana.f}: the parton-to-hadron routine calls a function \texttt{GET\_DELTA\_TAU(b)} that returns the sigmoid value, which is added to the coalescence formation time of every produced hadron. The two parton cross sections, $\sigma=0.3$ and $0.7$~mb, correspond to the two Debye masses $\mu$ quoted after Eq.~(\ref{Eq-sigma-zpc}), and the $(A,B,C)$ parameter sets quoted above are tuned independently for each.

\FloatBarrier
\section{Model Validation\label{sec:validation}}

\subsection{Bjorken energy density}

The first validation is the Bjorken energy density as a function of proper time $\tau$,
\be
\varepsilon_{Bj}(\tau) = \frac{1}{\tau\, A_{\perp}}\frac{dE_{T}}{dy}\bigg|_{y=0},
\label{Eq-eBj}
\ee
where $A_{\perp}$ is the transverse overlap area, computed for several impact-parameter bins.

\begin{figure}[!htbp]
\centering
\includegraphics[width=\linewidth,clip]{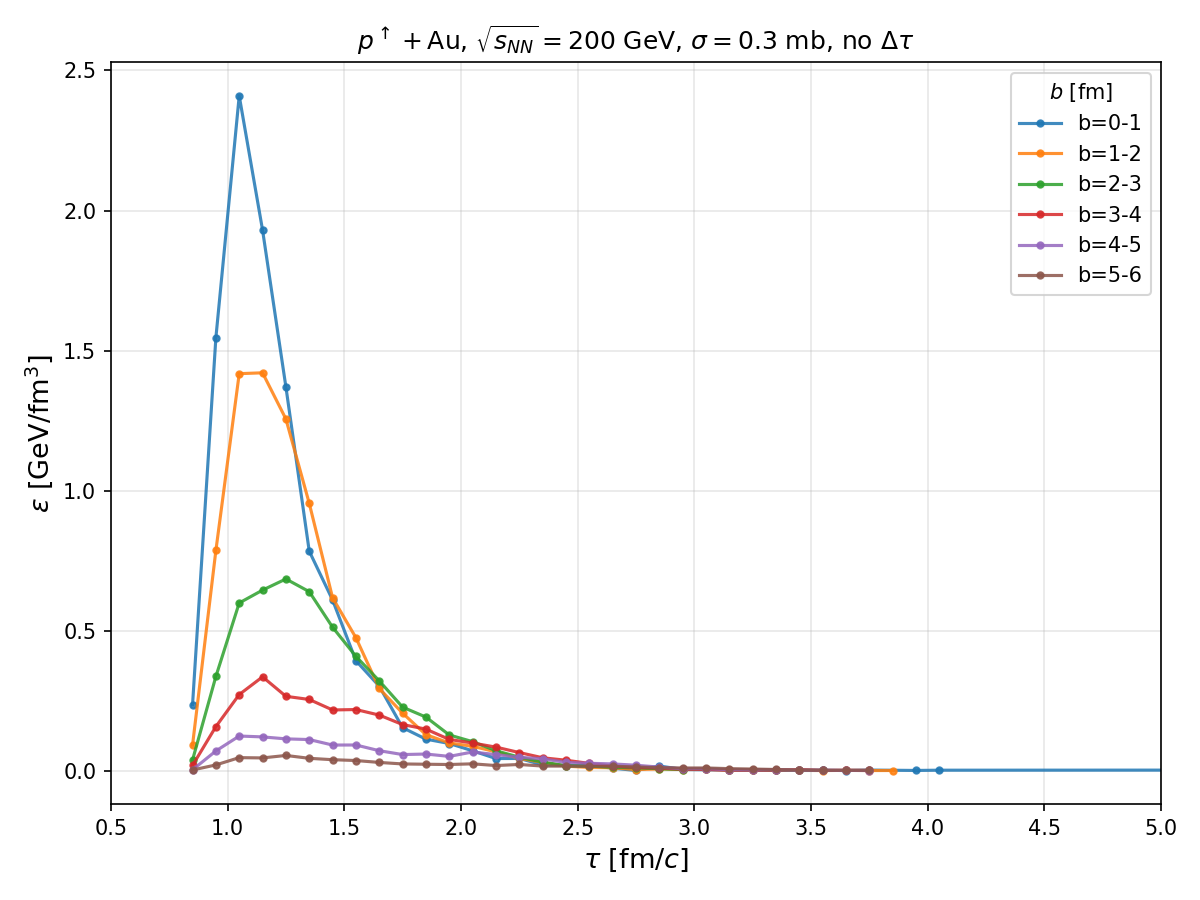}
\caption{Bjorken energy density $\varepsilon_{Bj}(\tau)$ in $p+$Au collisions at $\sqrt{s_{NN}}=200$~GeV without the formation-time delay, for $\sigma=0.3$~mb, shown for several impact-parameter bins.}
\label{fig:ed-B1}
\end{figure}

\begin{figure}[!htbp]
\centering
\includegraphics[width=\linewidth,clip]{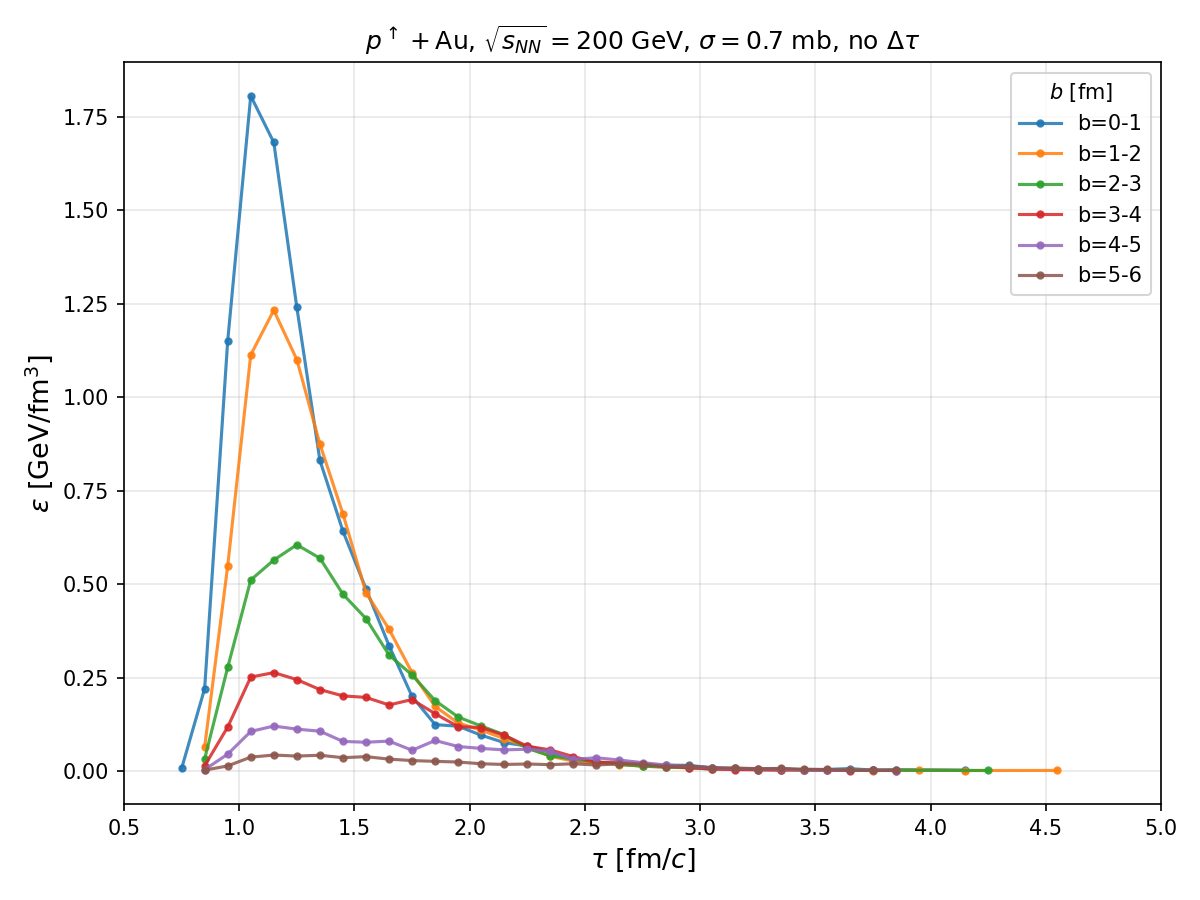}
\caption{Same as Fig.~\ref{fig:ed-B1} but for $\sigma=0.7$~mb.}
\label{fig:ed-B2}
\end{figure}

\begin{figure}[!htbp]
\centering
\includegraphics[width=\linewidth,clip]{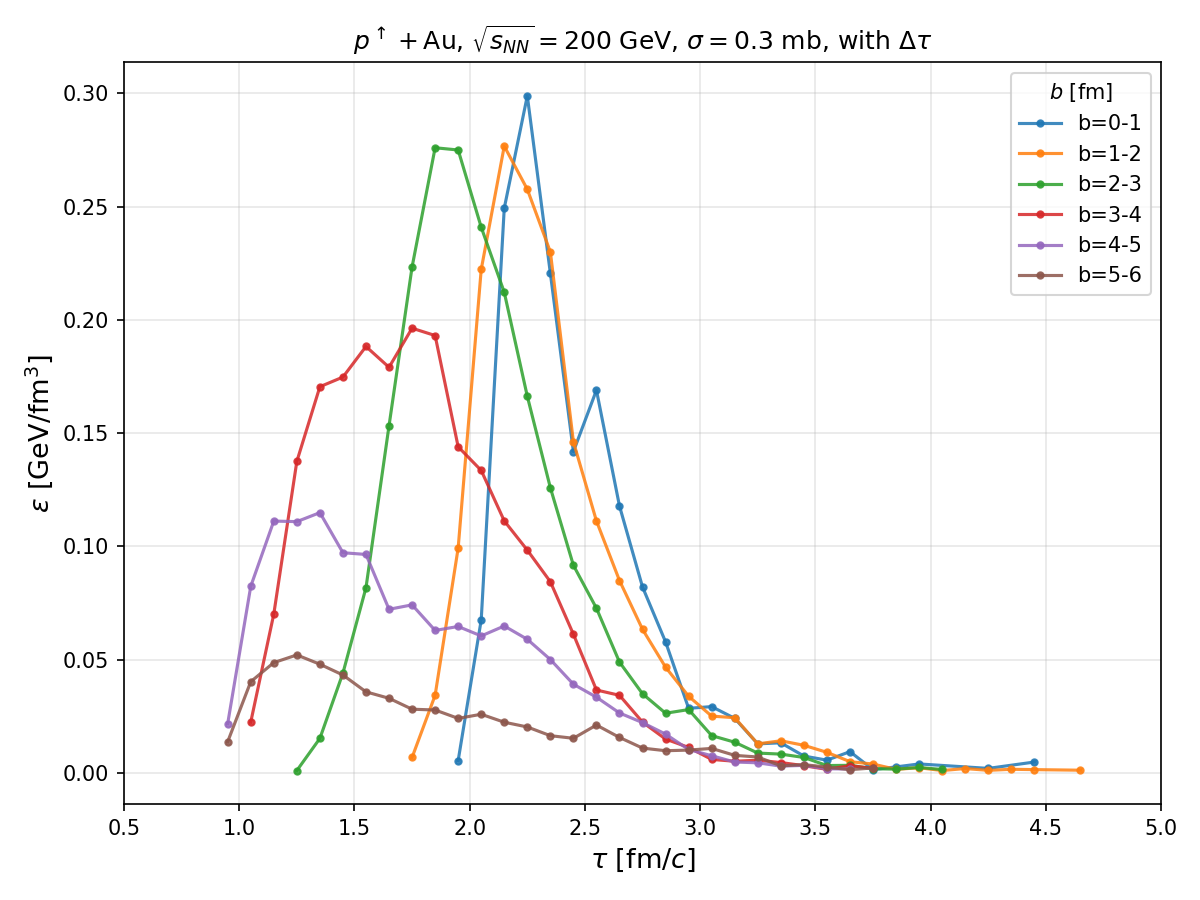}
\caption{Bjorken energy density $\varepsilon_{Bj}(\tau)$ in $p+$Au at $\sqrt{s_{NN}}=200$~GeV with the formation-time delay of Eq.~(\ref{Eq-dtau}), for $\sigma=0.3$~mb, shown for several impact-parameter bins. The peak central value is reduced to $\sim 0.3$~GeV/fm$^3$.}
\label{fig:ed-D1}
\end{figure}

\begin{figure}[!htbp]
\centering
\includegraphics[width=\linewidth,clip]{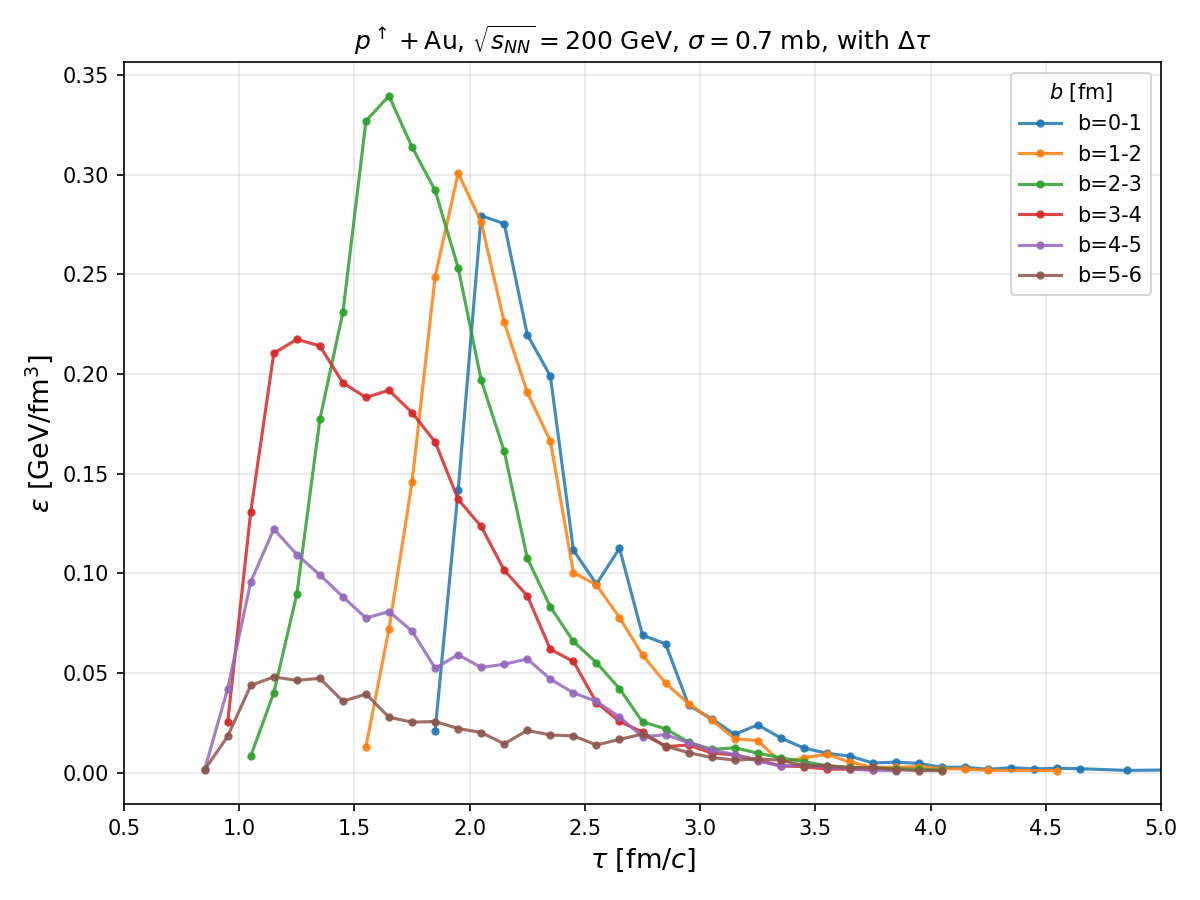}
\caption{Same as Fig.~\ref{fig:ed-D1} but for $\sigma=0.7$~mb.}
\label{fig:ed-D2}
\end{figure}

Figures~\ref{fig:ed-B1} and~\ref{fig:ed-B2} show $\varepsilon_{Bj}(\tau)$ without the formation-time delay. The most central bin ($b=0$--$1$~fm) reaches peak values of $\sim 2.4$ and $\sim 1.8$~GeV/fm$^3$ for $\sigma=0.3$ and $0.7$~mb respectively, which are unphysically high for $p+$Au. Turning on $\Delta\tau(b)$ with the parameters of Sec.~\ref{sec:ftime}, Figs.~\ref{fig:ed-D1} and~\ref{fig:ed-D2}, brings the peak central density down to $\sim 0.3$~GeV/fm$^3$, in line with the value expected for small systems~\cite{Zhao:2024lpc}, while leaving the shape and peripheral behaviour of $\varepsilon_{Bj}(\tau)$ qualitatively unchanged.

\subsection{Elliptic flow $v_2(p_T)$}

As a second, more stringent check we compute $v_2(p_T)$ for charged hadrons in 0--5\% $p+$Au using an event-plane procedure designed to mimic the recent PHENIX analysis~\cite{PHENIX:2018lia} . The centrality is determined from the charged-hadron multiplicity in the BBC south acceptance ($-3.9 < \eta < -3.1$). The event plane is reconstructed from charged hadrons in the FVTX south window ($-3.0 < \eta < -1.0$), and its resolution is estimated by the two-sub-event method using $-3.0 < \eta < -2.0$ and $-2.0 < \eta < -1.0$ as the sub-events. The $v_2$ is then measured for charged hadrons at midrapidity, $|\eta|<0.35$.

\begin{figure}[!htbp]
\centering
\includegraphics[width=\linewidth,clip]{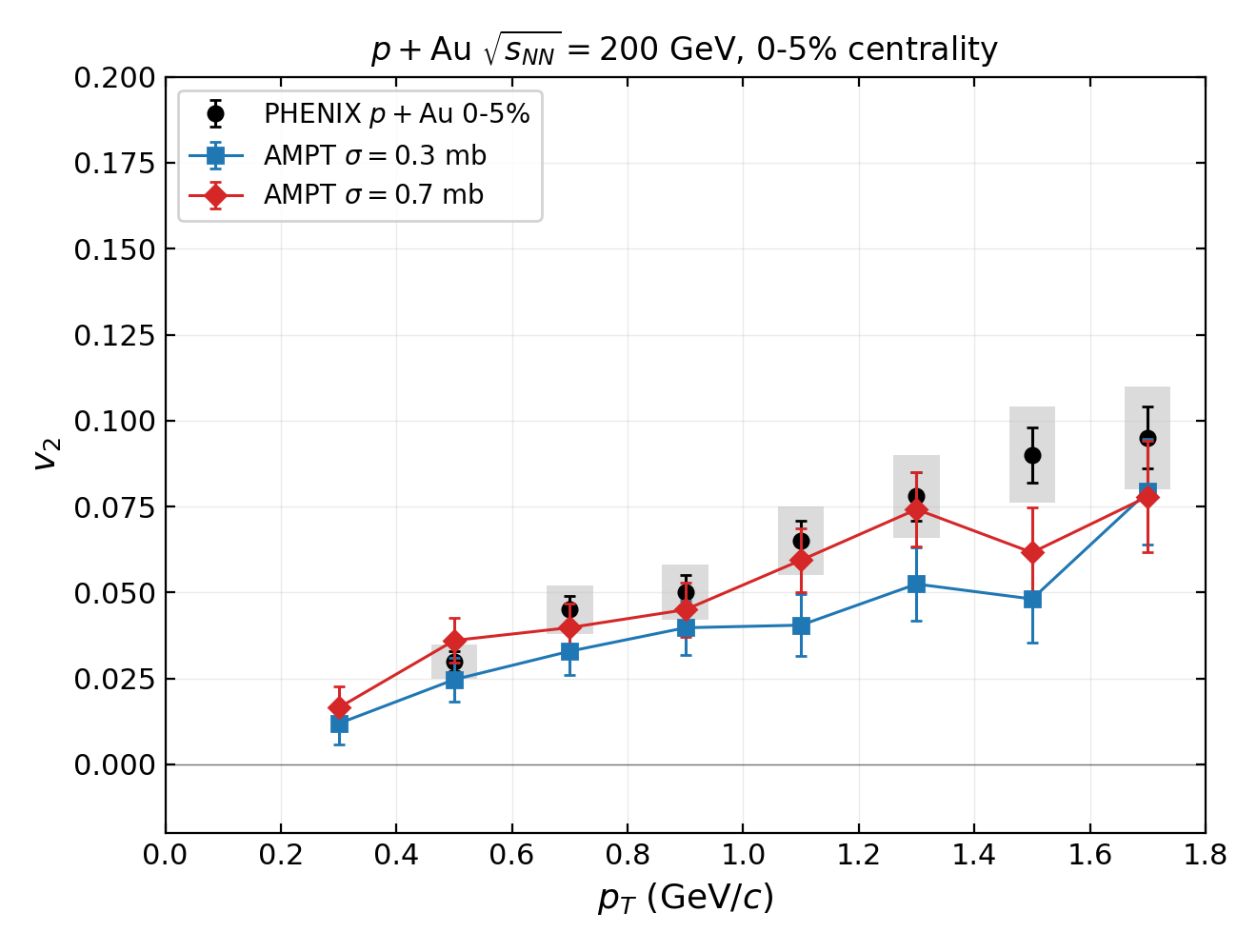}
\caption{Charged-hadron $v_2(p_T)$ in 0--5\% $p+$Au at $\sqrt{s_{NN}}=200$~GeV obtained from the improved AMPT model with the formation-time delay (red diamonds and blue squares) compared to the PHENIX data~\cite{PHENIX:2018lia} (black circles).}
\label{fig:v2}
\end{figure}

As shown in Fig.~\ref{fig:v2}, the model with the formation-time delay reproduces the PHENIX data well over $0.4 < p_T < 2.0$~GeV/$c$. The agreement is non-trivial: $\Delta\tau(b)$ was tuned only against the energy density, and the fact that it leaves $v_2(p_T)$ in good agreement with the measurement gives us confidence that the partonic and hadronic phases are reasonably well described by the present setup.

\FloatBarrier
\section{Results of Charge Separation\label{sec:results}}

We now present the main results: the three-point correlator $\gamma_{\alpha\beta}$ for $p^{\uparrow}+$Au at $b=1.5$~fm, in the four geometry schemes and for the two parton cross sections, shown at three stages of the evolution: (i) the initial parton level, immediately after the event-by-event $p_y$ exchange of Sec.~\ref{sec:cme} but before the parton cascade; (ii) the after-ZPC parton level, i.e. after the parton cascade but before quark coalescence; and (iii) the final state, after quark coalescence and the ART hadronic rescattering. In each panel we display $\gamma_{OS}$ and $\gamma_{SS}$ separately, so that both the overall correlator and the charge-dependent difference $\Delta\gamma=\gamma_{OS}-\gamma_{SS}$ can be read off directly.

\subsection{$P_+$ dependence}

We firstly examine $\gamma_{\alpha\beta}$ as a function of the pair transverse momentum $P_+ = (p_{T,\alpha}+p_{T,\beta})/2$.

\begin{figure}[!htbp]
\centering
\includegraphics[width=\linewidth,clip]{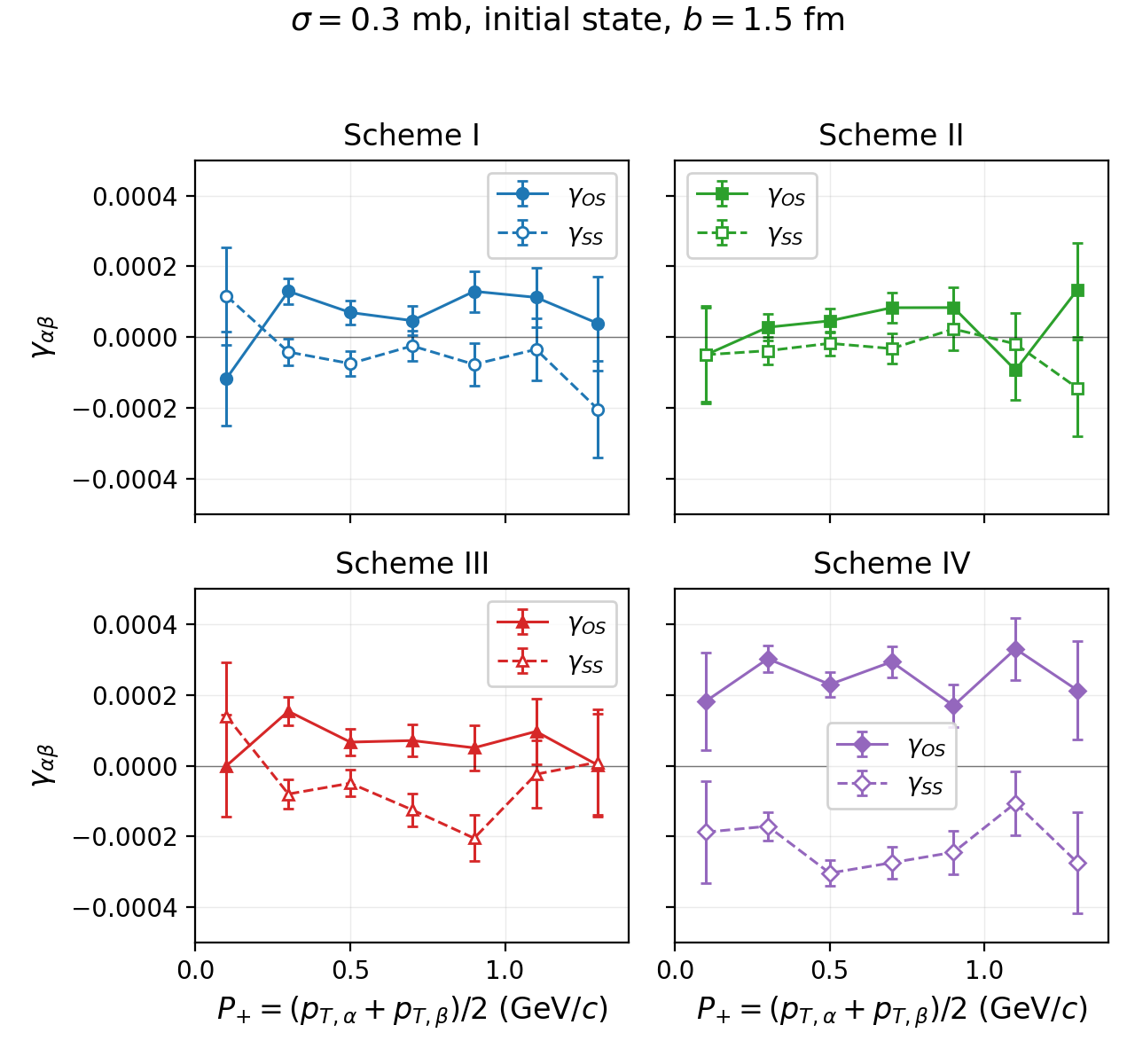}
\caption{Initial-state $\gamma_{OS}$ and $\gamma_{SS}$ versus $P_+$ for the four geometry schemes in $p^{\uparrow}+$Au at $b=1.5$~fm with $\sigma=0.3$~mb. All four schemes show a clear charge-dependent splitting; scheme IV is the largest, scheme II the smallest.}
\label{fig:gamma-pplus-ini-03}
\end{figure}

\begin{figure}[!htbp]
\centering
\includegraphics[width=\linewidth,clip]{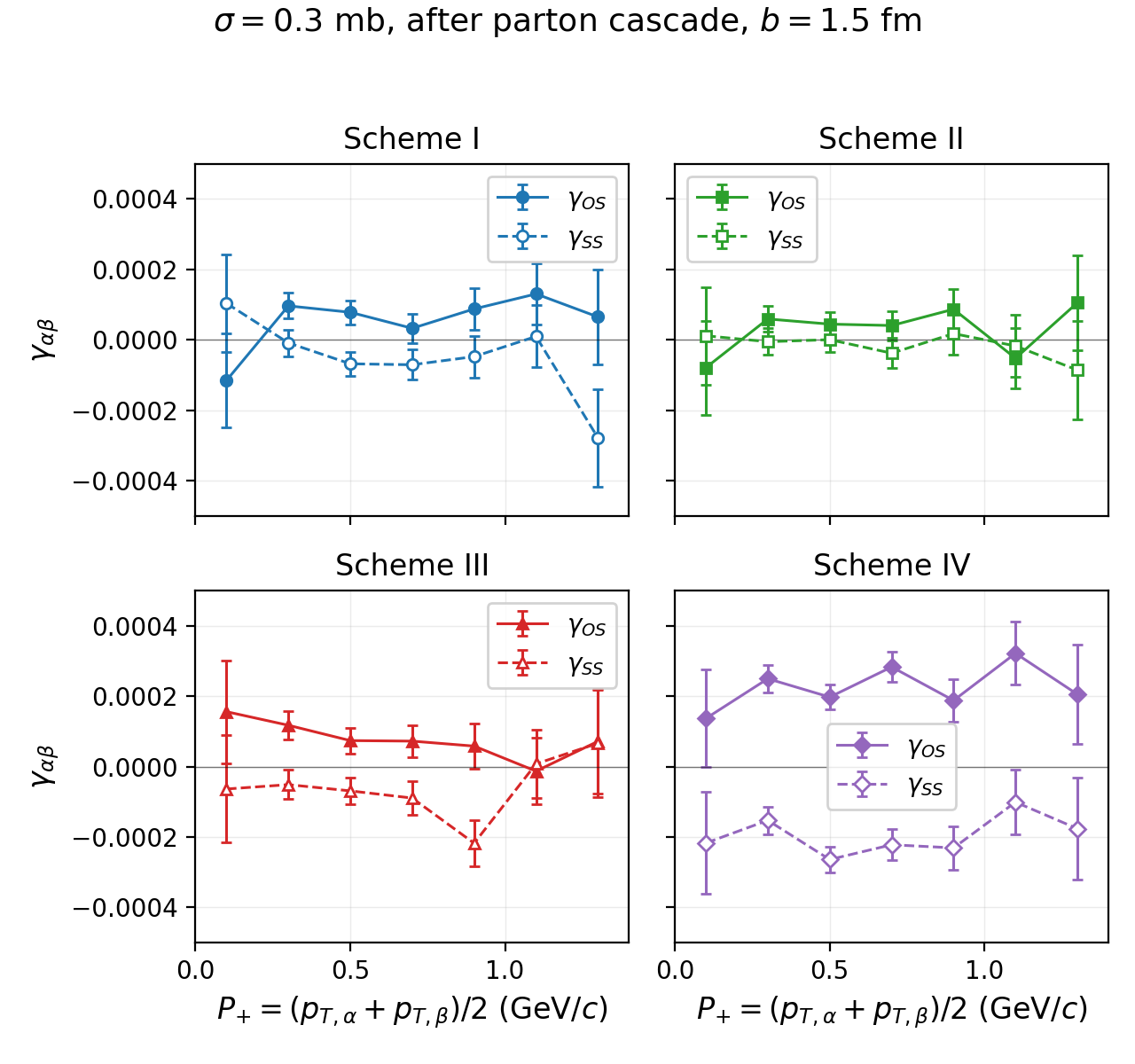}
\caption{Same as Fig.~\ref{fig:gamma-pplus-ini-03} but after the parton cascade (ZPC), before quark coalescence. The signal at this stage is essentially identical to the initial state, indicating that the parton cascade dissipates only $\sim 10$--$20\%$ of $|\Delta\gamma|$.}
\label{fig:gamma-pplus-zpc-03}
\end{figure}

\begin{figure}[!htbp]
\centering
\includegraphics[width=\linewidth,clip]{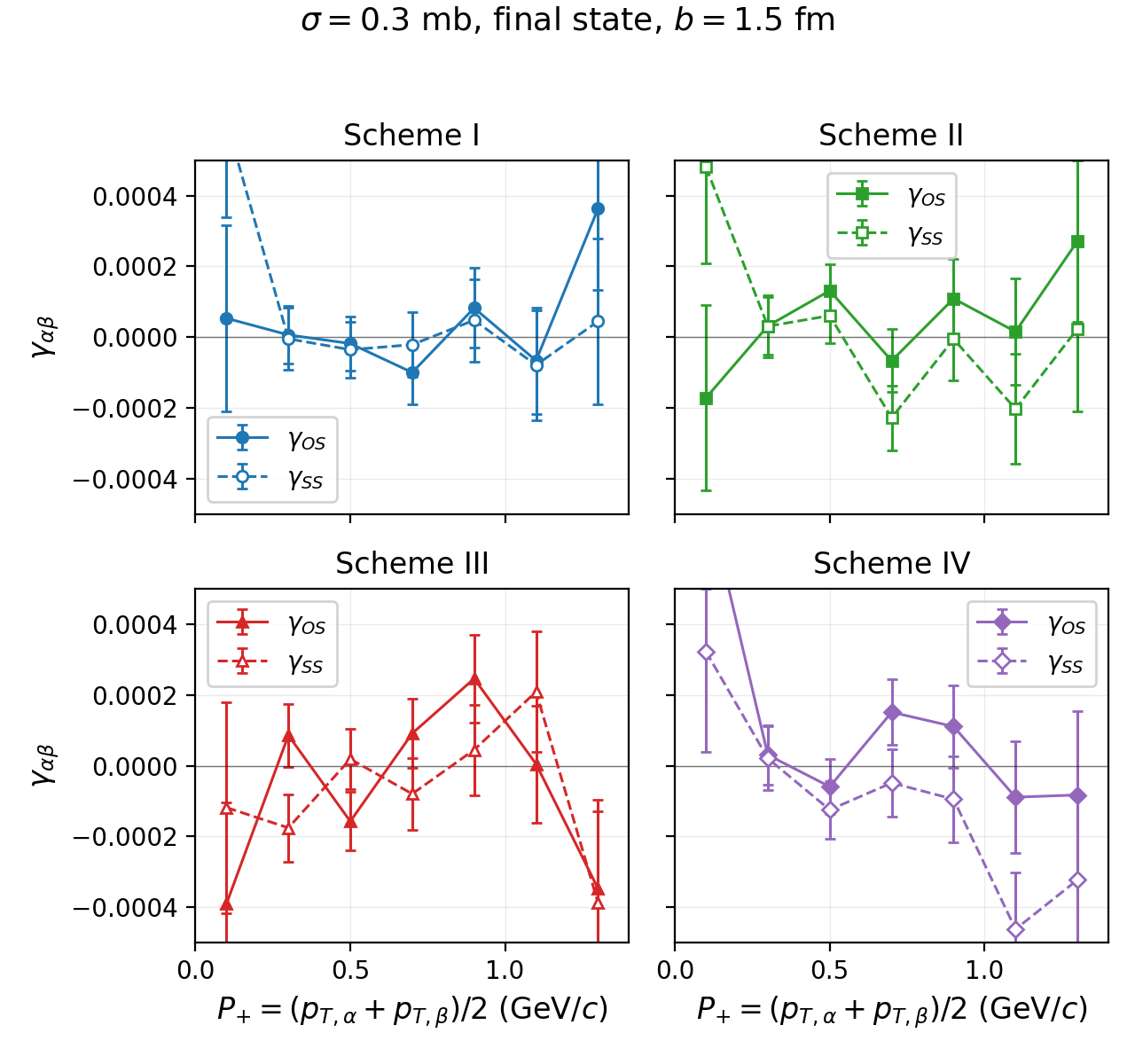}
\caption{Same as Fig.~\ref{fig:gamma-pplus-ini-03} but after full parton cascade and hadronic rescattering (ART). Most of the $|\Delta\gamma|$ dissipation occurs in this stage; scheme~IV continues to display a clearly visible splitting between $\gamma_{OS}$ and $\gamma_{SS}$.}
\label{fig:gamma-pplus-fin-03}
\end{figure}

\begin{figure}[!htbp]
\centering
\includegraphics[width=\linewidth,clip]{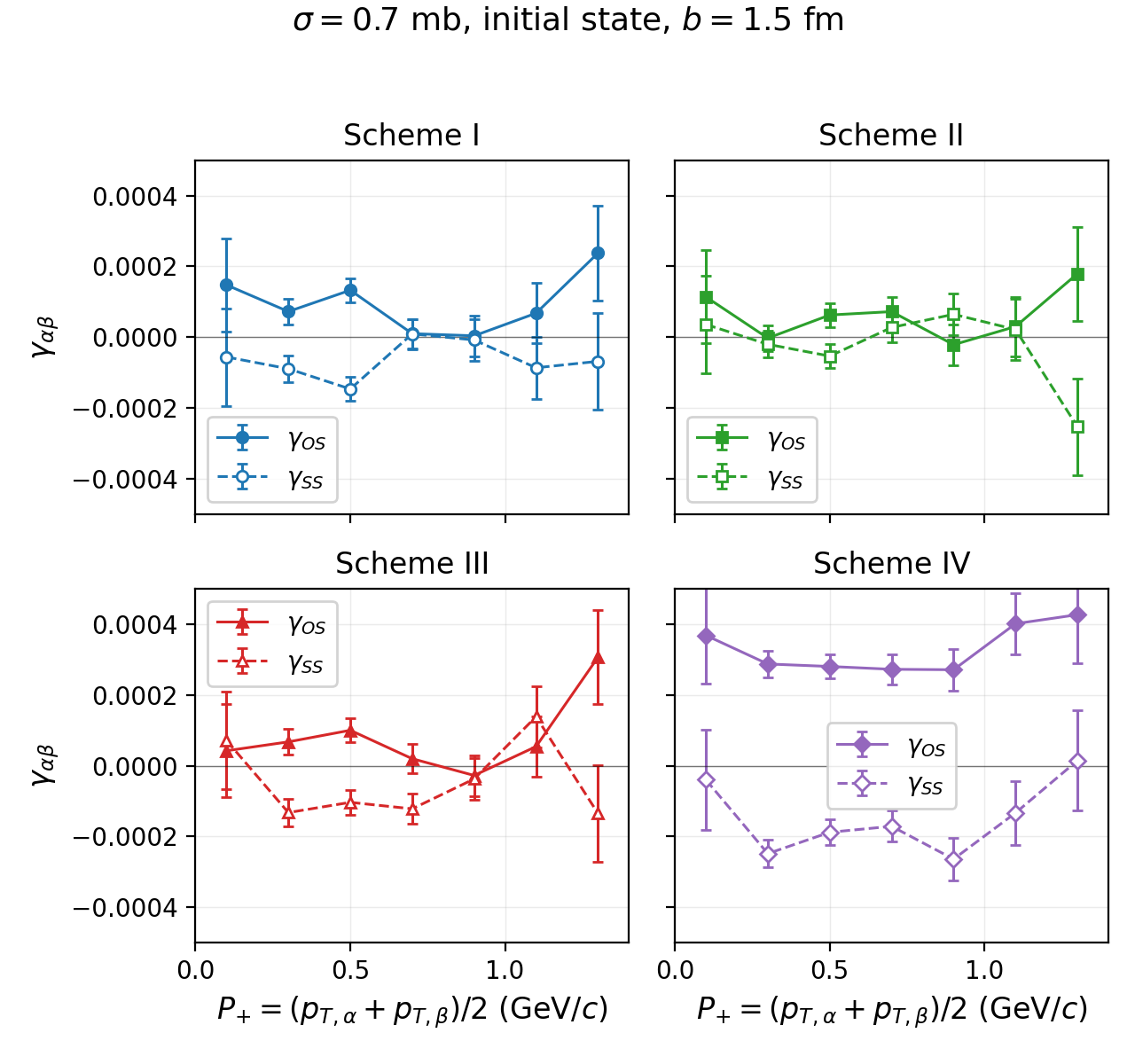}
\caption{Same as Fig.~\ref{fig:gamma-pplus-ini-03} but for $\sigma=0.7$~mb.}
\label{fig:gamma-pplus-ini-07}
\end{figure}

\begin{figure}[!htbp]
\centering
\includegraphics[width=\linewidth,clip]{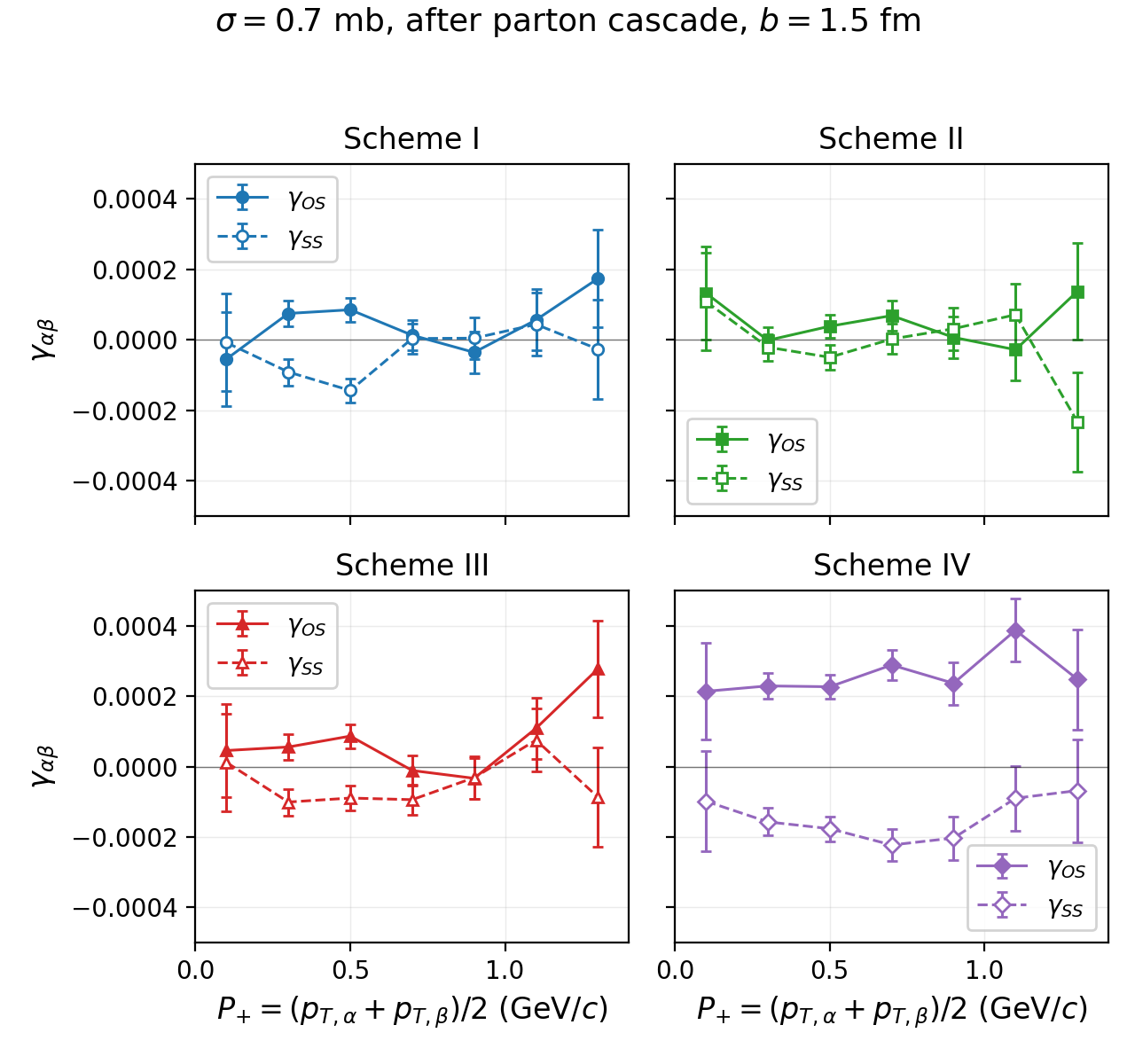}
\caption{Same as Fig.~\ref{fig:gamma-pplus-zpc-03} but for $\sigma=0.7$~mb.}
\label{fig:gamma-pplus-zpc-07}
\end{figure}

\begin{figure}[!htbp]
\centering
\includegraphics[width=\linewidth,clip]{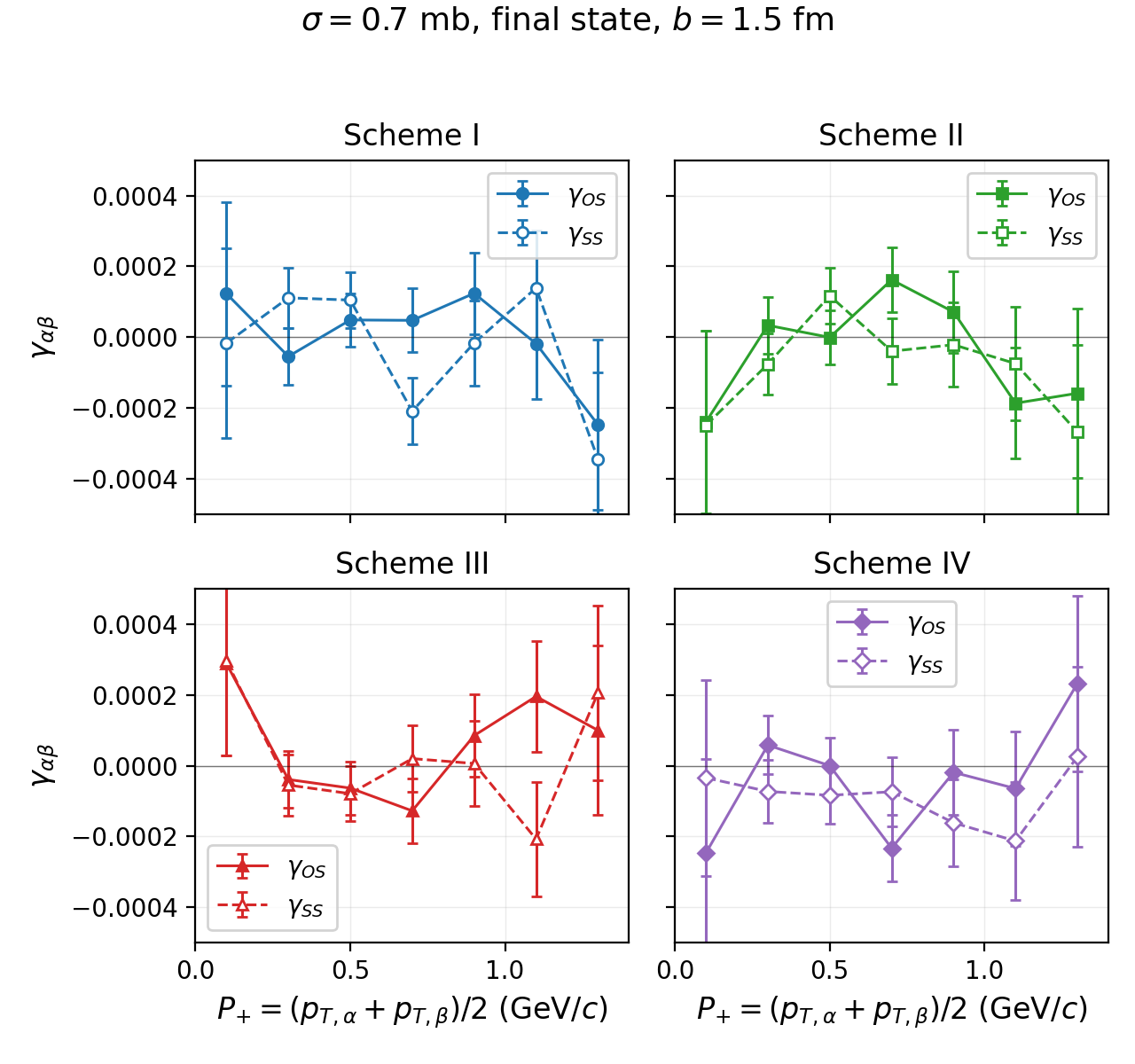}
\caption{Same as Fig.~\ref{fig:gamma-pplus-fin-03} but for $\sigma=0.7$~mb.}
\label{fig:gamma-pplus-fin-07}
\end{figure}

At the initial state (Figs.~\ref{fig:gamma-pplus-ini-03} and~\ref{fig:gamma-pplus-ini-07} corresponding to $\sigma=0.3$~mb and $\sigma=0.7$~mb) all four schemes display a clear splitting between opposite- and same-sign pairs, with a hierarchy of
\begin{equation*}
|\Delta\gamma|_{\mathrm{IV}} \gg |\Delta\gamma|_{\mathrm{I}} \approx |\Delta\gamma|_{\mathrm{III}} > |\Delta\gamma|_{\mathrm{II}}.
\end{equation*}
This ordering reproduces the scheme-by-scheme hierarchy of $B^2$ at the overlap centre that we computed in Ref.~\cite{Wu:2024vcd}: in scheme IV the impact parameter points along $-y$, the direction in which the charge asymmetry of the polarized proton constructively enhances $B^2$, while scheme II picks up the opposite, destructive geometry. Quantitatively, averaging the $P_+$ projection over $0.2 \le P_+ \le 1.0$~GeV/$c$, scheme~IV yields $|\Delta\gamma| \simeq 5\times 10^{-4}$, scheme~I and III about $1.5\times 10^{-4}$, and scheme~II about $0.5$--$0.8\times 10^{-4}$, with the same pattern observed in the $\Delta\eta$ projection.

After full evolution (Figs.~\ref{fig:gamma-pplus-fin-03} and~\ref{fig:gamma-pplus-fin-07}) the parton cascade and the hadronic rescattering together dissipate most of the initial splitting. The decomposition into the two stages is informative: comparing the after-ZPC parton-level correlator (Figs.~\ref{fig:gamma-pplus-zpc-03} and~\ref{fig:gamma-pplus-zpc-07}) with the initial parton-level one (Figs.~\ref{fig:gamma-pplus-ini-03} and~\ref{fig:gamma-pplus-ini-07}), we find that ZPC alone reduces $|\Delta\gamma|$ by only $\sim 10$--$20\%$, whereas the subsequent hadron-coalescence and ART hadronic rescattering bring it down to $\sim 10$--$30\%$ of the initial level. The bulk of the dissipation therefore happens in the hadronic phase rather than during the parton cascade. Despite this large absolute reduction, scheme~IV continues to display a clearly visible gap between $\gamma_{OS}$ (positive) and $\gamma_{SS}$ (negative) across the full $P_+$ range probed, and the central-value hierarchy $|\Delta\gamma|_{\mathrm{IV}} > |\Delta\gamma|_{\mathrm{I,III}} > |\Delta\gamma|_{\mathrm{II}}$ is preserved at both cross sections, so that the inter-scheme difference $\Delta\gamma_{\mathrm{IV}}-\Delta\gamma_{\mathrm{II}}$ remains a non-trivial observable. The much smaller survival fraction quoted here, compared with the $\sim 40\%$ figure of Ref.~\cite{Xu:2025cme}, originates in the much smaller parton cross sections used in this work ($0.3$ and $0.7$~mb instead of $10$~mb), which shorten the parton-cascade phase and thereby leave a longer effective window for the ART hadronic rescattering to dissipate the charge separation.

\subsection{$\Delta\eta$ dependence}

The pseudorapidity-gap dependence of $\gamma_{\alpha\beta}$ provides a complementary handle, because the QCD background associated with $v_2$ is expected to fall off rapidly with $\Delta\eta = |\eta_\alpha - \eta_\beta|$ while the CME signal is long-range in $\eta$.

\begin{figure}[!htbp]
\centering
\includegraphics[width=\linewidth,clip]{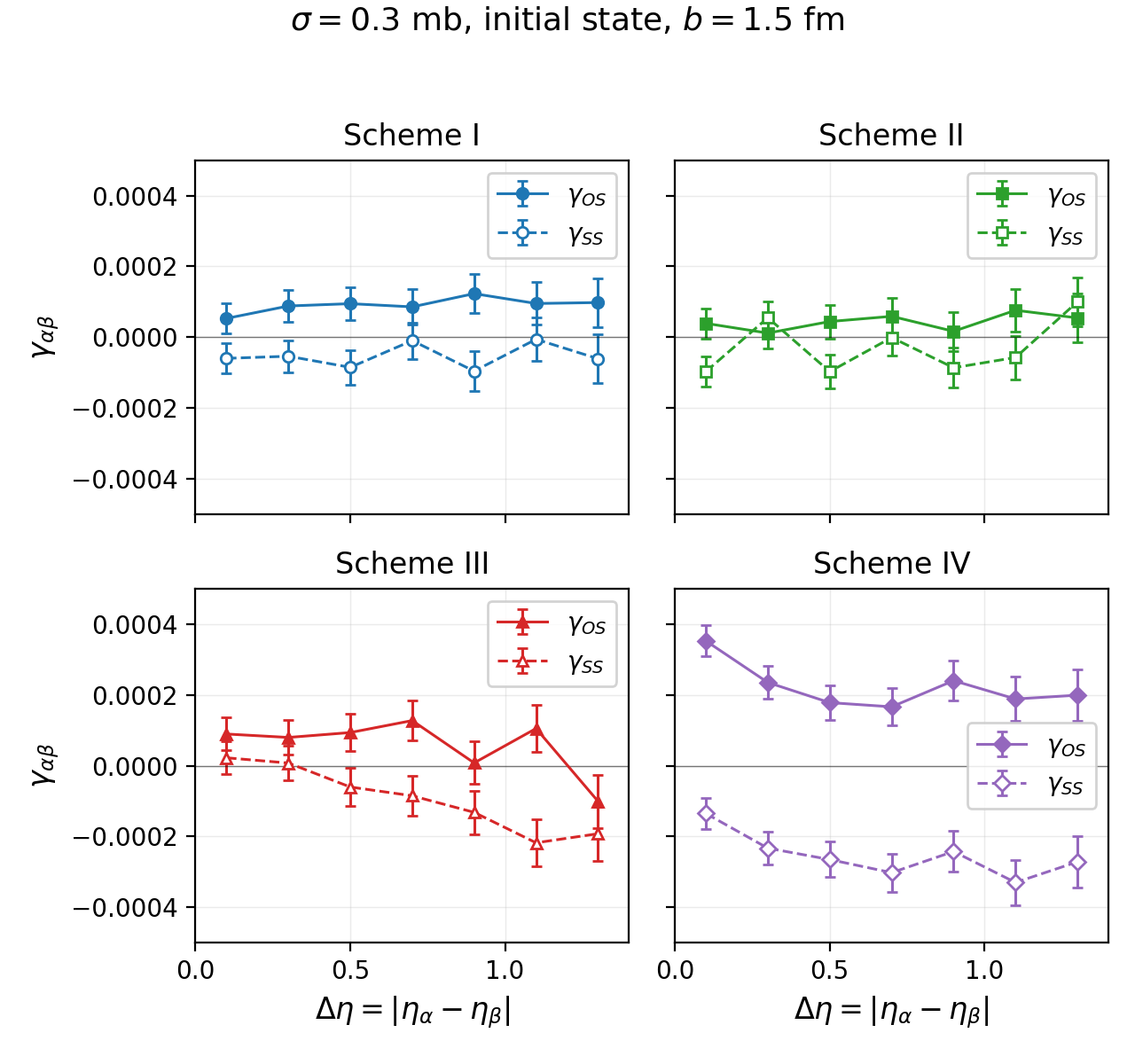}
\caption{Initial-state $\gamma_{OS}$ and $\gamma_{SS}$ versus $\Delta\eta$ for the four geometry schemes at $b=1.5$~fm with $\sigma=0.3$~mb.}
\label{fig:gamma-deta-ini-03}
\end{figure}

\begin{figure}[!htbp]
\centering
\includegraphics[width=\linewidth,clip]{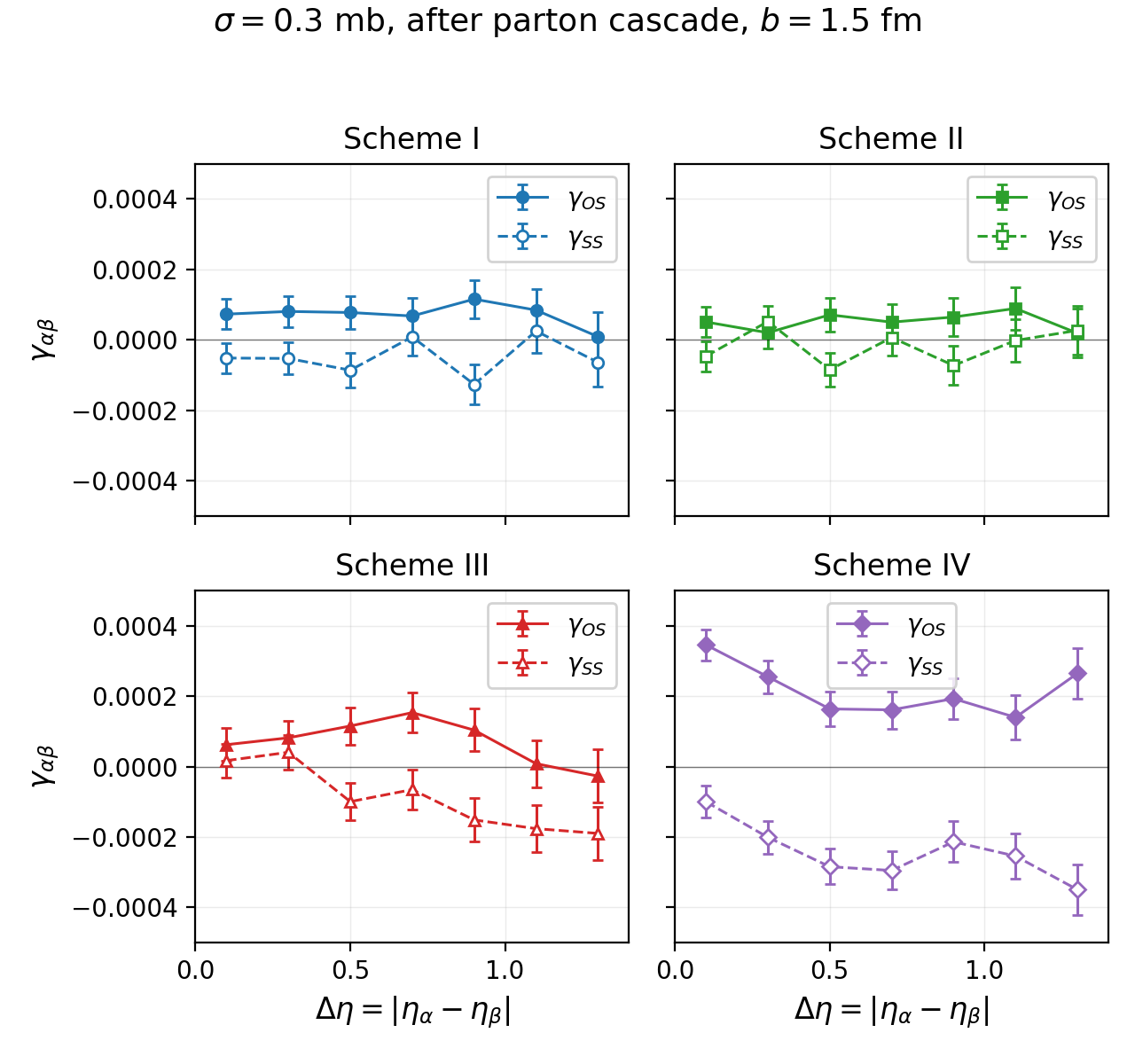}
\caption{Same as Fig.~\ref{fig:gamma-deta-ini-03} but after the parton cascade (ZPC), before quark coalescence.}
\label{fig:gamma-deta-zpc-03}
\end{figure}

\begin{figure}[!htbp]
\centering
\includegraphics[width=\linewidth,clip]{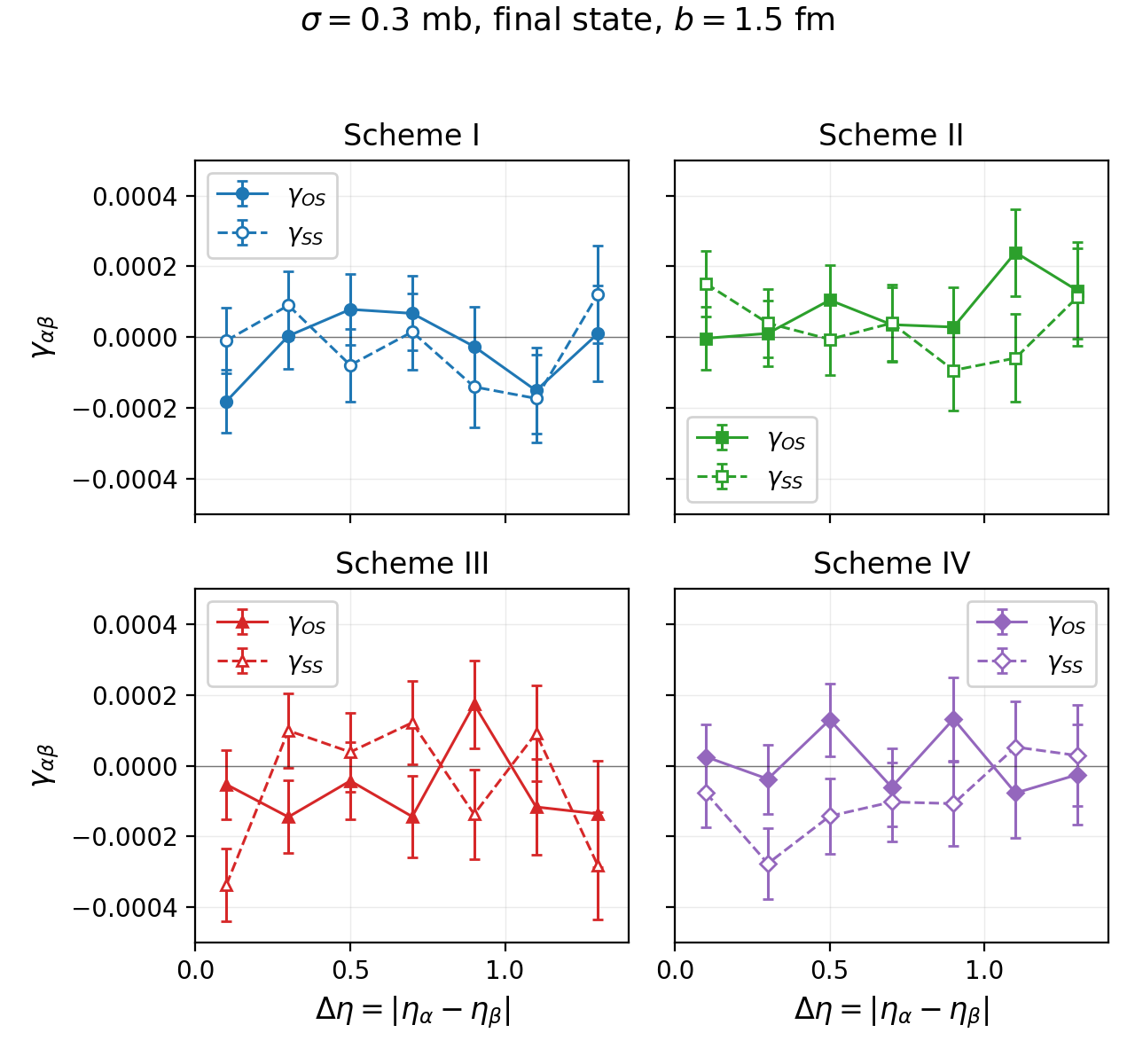}
\caption{Same as Fig.~\ref{fig:gamma-deta-ini-03} but after full parton cascade and hadronic rescattering (ART).}
\label{fig:gamma-deta-fin-03}
\end{figure}

\begin{figure}[!htbp]
\centering
\includegraphics[width=\linewidth,clip]{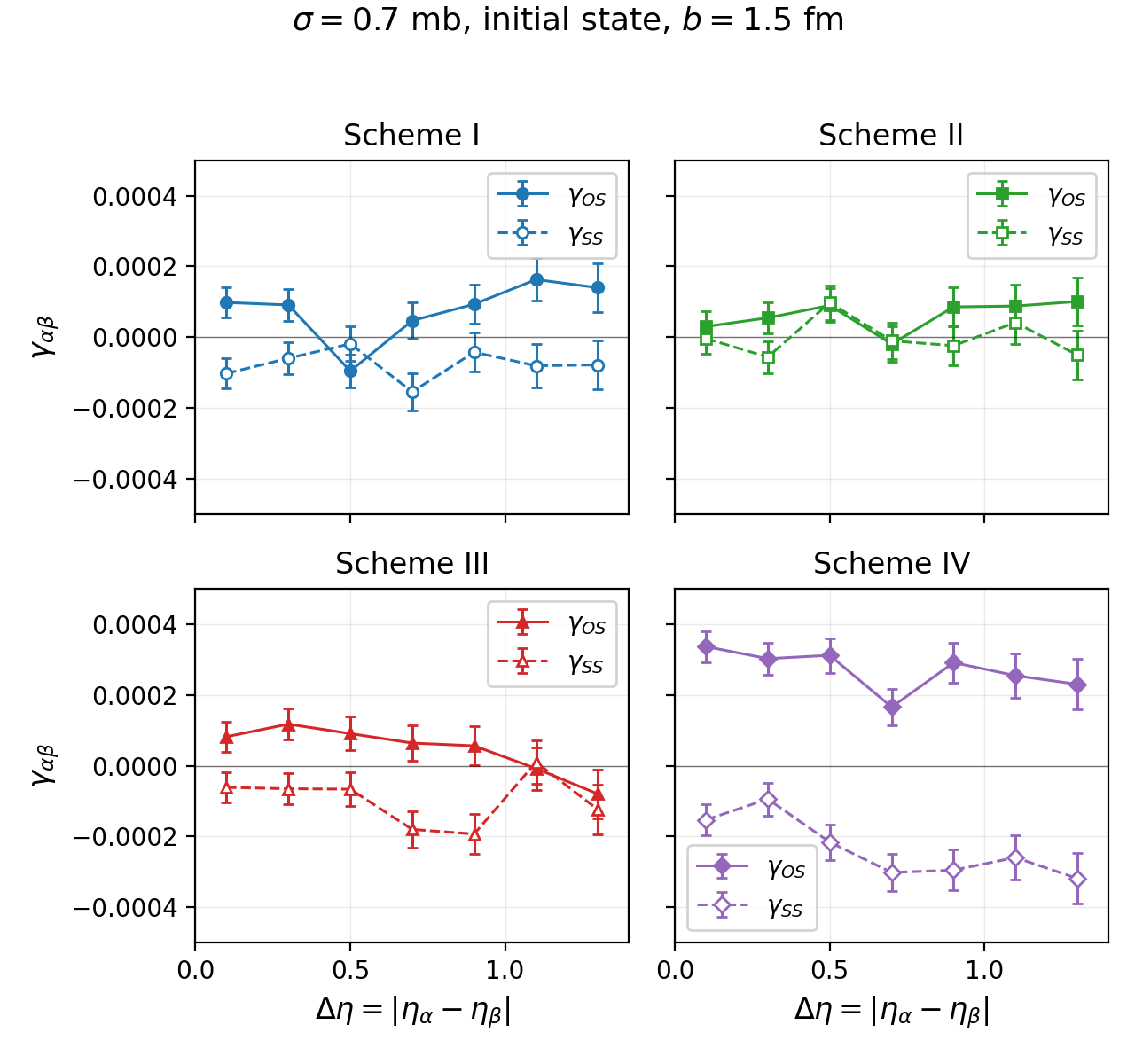}
\caption{Same as Fig.~\ref{fig:gamma-deta-ini-03} but for $\sigma=0.7$~mb.}
\label{fig:gamma-deta-ini-07}
\end{figure}

\begin{figure}[!htbp]
\centering
\includegraphics[width=\linewidth,clip]{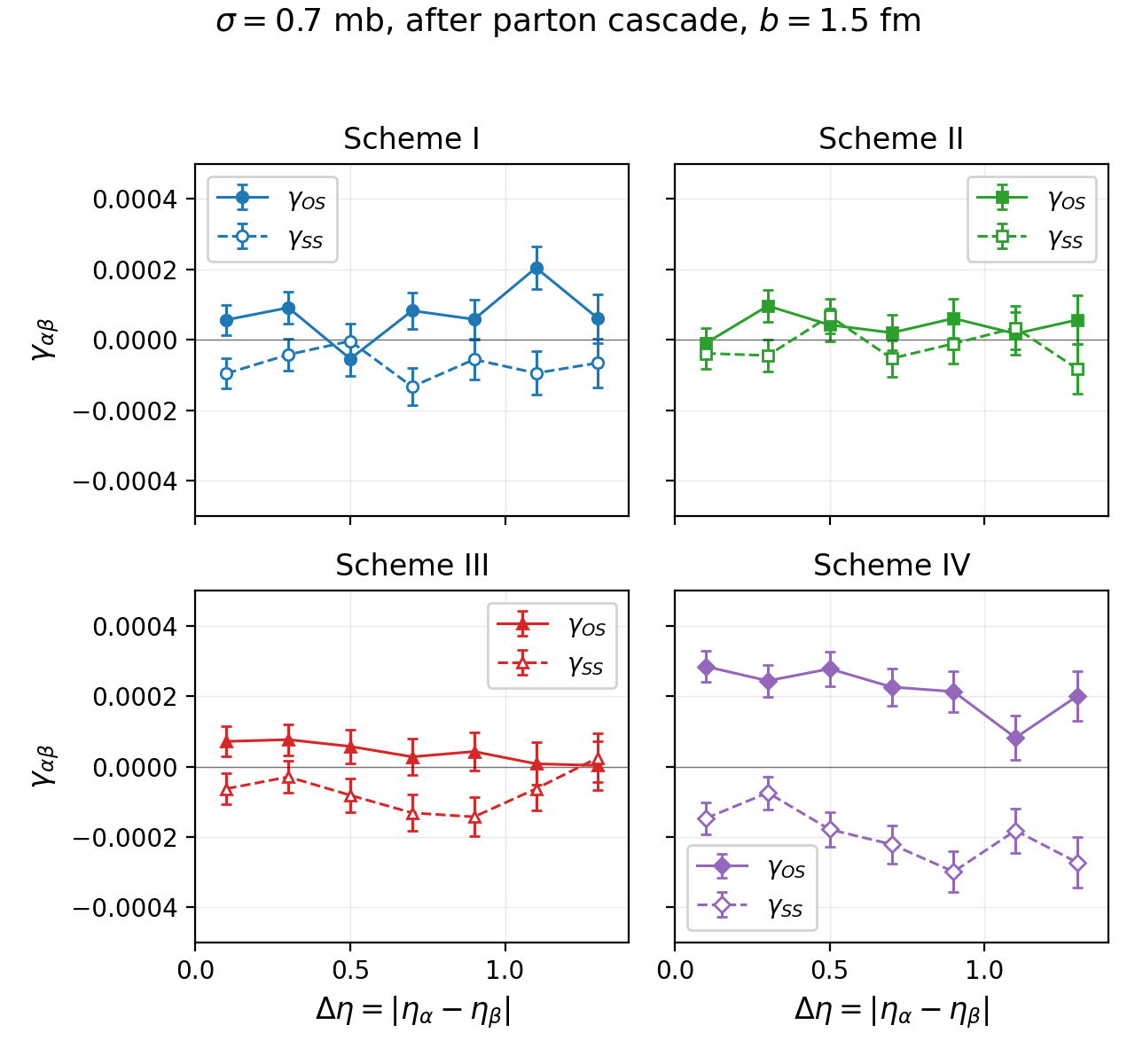}
\caption{Same as Fig.~\ref{fig:gamma-deta-zpc-03} but for $\sigma=0.7$~mb.}
\label{fig:gamma-deta-zpc-07}
\end{figure}

\begin{figure}[!htbp]
\centering
\includegraphics[width=\linewidth,clip]{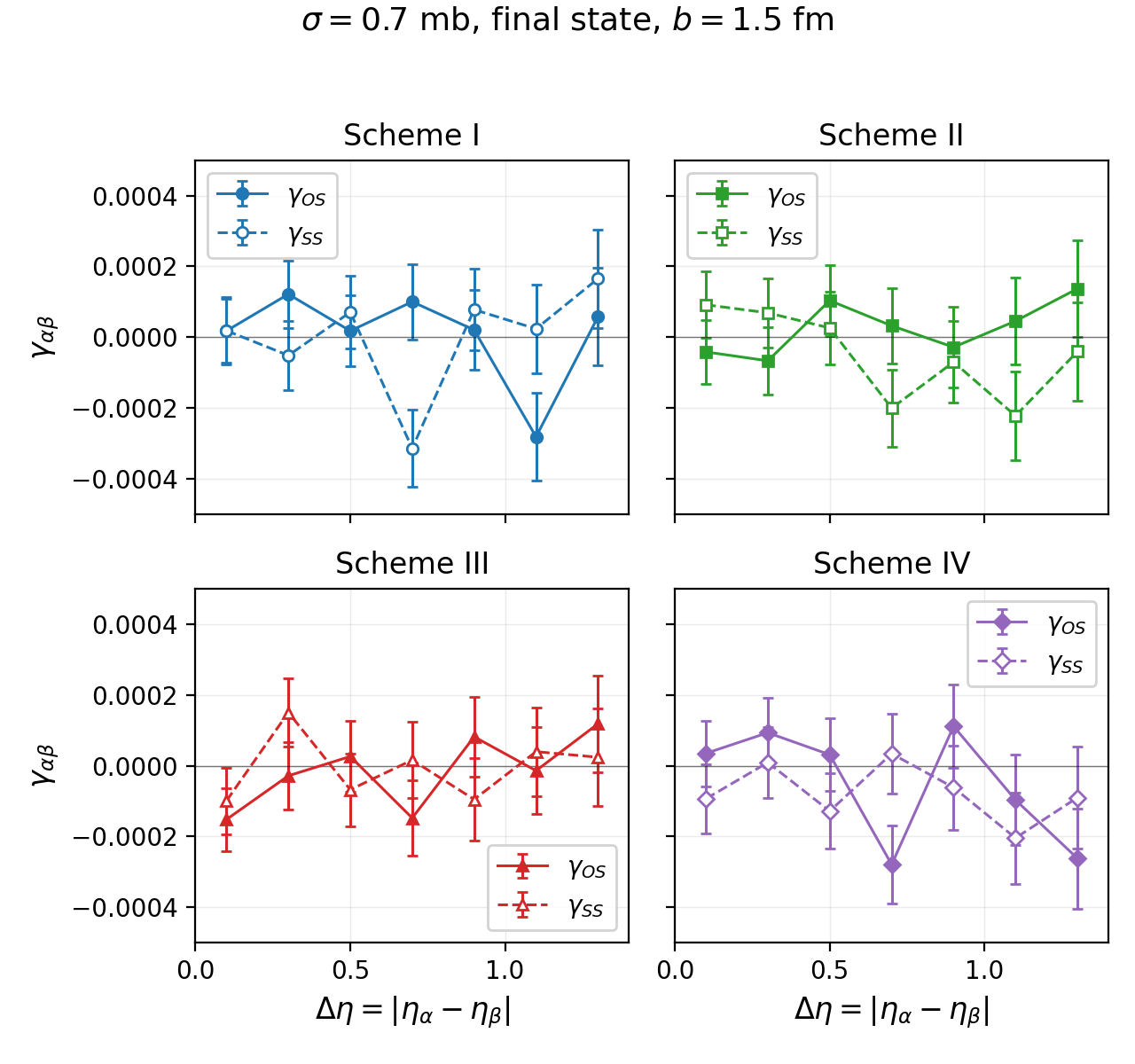}
\caption{Same as Fig.~\ref{fig:gamma-deta-fin-03} but for $\sigma=0.7$~mb.}
\label{fig:gamma-deta-fin-07}
\end{figure}

At the initial level (Figs.~\ref{fig:gamma-deta-ini-03} and~\ref{fig:gamma-deta-ini-07}) the splitting is essentially flat in $\Delta\eta$, which reflects the long-range nature of the source: the $p_y$ exchange is applied uniformly along the longitudinal direction, so the resulting correlator carries no intrinsic $\eta$ scale. The same scheme hierarchy as in the $P_+$ projection appears here. The after-ZPC stage (Figs.~\ref{fig:gamma-deta-zpc-03} and~\ref{fig:gamma-deta-zpc-07}) preserves both the magnitude and the flat profile of the initial state, consistent with the small ZPC-stage suppression observed in the $P_+$ projection. After the full evolution (Figs.~\ref{fig:gamma-deta-fin-03} and~\ref{fig:gamma-deta-fin-07}) the magnitude shrinks substantially, but a clearly visible $\gamma_{OS}/\gamma_{SS}$ gap survives in scheme~IV at both $\sigma=0.3$ and $0.7$~mb, with $\gamma_{OS}$ remaining positive and $\gamma_{SS}$ negative across the $\Delta\eta$ range, and a profile that remains broadly flat. In schemes I, II, and III the residual splitting is smaller, but the central-value hierarchy is the same as in the initial state. The persistence of a flat profile under evolution distinguishes $p^{\uparrow}+$Au from Au+Au, where the strong short-range $\Delta\eta$ component of the correlator is dominated by background~\cite{Ma:2011uma}.

\subsection{Discussion}

Two physics messages emerge. First, the event-by-event implementation of Sec.~\ref{sec:cme} automatically inherits the geometric hierarchy of $B^2$ across the four schemes, and so produces an initial-state correlator that already encodes the geometric signature of the CME. The same hierarchy persists, although with a much reduced amplitude, through parton cascade and hadronic rescattering. Second, because the background contribution to $\Delta\gamma$ in $p+$Au is small~\cite{Xu:2025cme}, the difference
\be
\Delta\gamma_{\mathrm{IV}} - \Delta\gamma_{\mathrm{II}} \;\approx\; \Delta\gamma^{\mathrm{CME}}_{\mathrm{IV}} - \Delta\gamma^{\mathrm{CME}}_{\mathrm{II}}
\label{Eq-scheme-diff}
\ee
isolates an essentially pure CME observable. Our results show that this difference remains clearly non-zero after full evolution, with $\Delta\gamma_{\mathrm{IV}}-\Delta\gamma_{\mathrm{II}}$ of order $1$--$2\times 10^{-4}$, which suggests that an experimental measurement with the Zero Degree Calorimeters tagging the impact-parameter orientation should be sensitive to it.

The two cross sections ($\sigma=0.3$ and $0.7$~mb) give qualitatively the same picture, both at the initial level and after evolution. This robustness against a factor-of-two variation in $\sigma$, combined with the validation against the PHENIX $v_2(p_T)$ data, supports the conclusion that the scheme-dependent pattern of $\Delta\gamma$ is dictated by the geometric structure of the magnetic field rather than by the details of the partonic transport.

\FloatBarrier
\section{Summary\label{sec:summary}}

We have studied the CME-induced charge separation in $p^{\uparrow}+$Au collisions at $\sqrt{s_{NN}}=200$~GeV with an improved string-melting AMPT model that incorporates two new ingredients.

First, the impact-parameter-dependent hadron formation-time delay of Eq.~(\ref{Eq-dtau}) brings the peak Bjorken energy density from the unphysical $\sim 2$~GeV/fm$^3$ produced by the default AMPT setup down to $\sim 0.3$~GeV/fm$^3$, and at the same time yields a $v_2(p_T)$ that agrees with the PHENIX 0--5\% $p+$Au measurement. Second, the event-by-event CME source of Eq.~(\ref{Eq-fraction}), in which the quark momentum-exchange fraction scales as $|\bB|_{\mathrm{event}}/|\bB|_{\mathrm{max}}$ and is capped at the 7\% value that fits the Au+Au $\gamma_{SS}$ correlator, replaces the fixed-fraction prescription used in our previous studies and so transmits the geometric structure of the magnetic field directly to the charge-separation observable.

With these two improvements we find that:
\begin{itemize}
\item the initial-state correlator shows a clear scheme hierarchy, $|\Delta\gamma|_{\mathrm{IV}} \gg |\Delta\gamma|_{\mathrm{I}} \approx |\Delta\gamma|_{\mathrm{III}} > |\Delta\gamma|_{\mathrm{II}}$, which tracks $B^2$ at the overlap centre, with $|\Delta\gamma|_{\mathrm{IV}}$ a factor of $\sim 5$--$10$ larger than $|\Delta\gamma|_{\mathrm{II}}$;
\item the parton cascade alone preserves $\sim 80$--$90\%$ of the initial signal; the bulk of the dissipation occurs in the hadronic phase (coalescence and ART rescattering), which brings the final-state $|\Delta\gamma|$ down to $\sim 10$--$30\%$ of its initial value;
\item after full evolution scheme~IV continues to display a clearly visible $\gamma_{OS}/\gamma_{SS}$ splitting in both the $P_+$ and $\Delta\eta$ projections, at both $\sigma=0.3$ and $0.7$~mb, and the central-value hierarchy $|\Delta\gamma|_{\mathrm{IV}} > |\Delta\gamma|_{\mathrm{I,III}} > |\Delta\gamma|_{\mathrm{II}}$ is preserved;
\item the correlator is essentially flat in $\Delta\eta$ both at the initial level and after evolution, which is consistent with the long-range character of the CME source in a small system;
\item the picture is robust against varying the parton cross section between $\sigma=0.3$ and $0.7$~mb.
\end{itemize}

The scheme-by-scheme pattern of $\Delta\gamma$, in particular the difference $\Delta\gamma_{\mathrm{IV}}-\Delta\gamma_{\mathrm{II}}$, provides an essentially background-free CME observable in $p^{\uparrow}+$Au. Our results suggest that a measurement using the Zero Degree Calorimeters to tag the impact-parameter orientation should be capable of resolving this difference after full QGP evolution, and so supports our broader proposal of using polarized proton--nucleus collisions to isolate the CME from $v_2$-related backgrounds.

Two aspects of the present analysis point directly to the limitations of the transport treatment used here. By construction the initial-state $\Delta\gamma$ is set by the $|\bB|$-scaled quark-momentum exchange applied before the parton cascade, and neither $\sigma$ nor $\Delta\tau(b)$ enters that step; consistently with this we find that the initial-state and after-ZPC $\gamma$ correlators are statistically indistinguishable between the $\sigma=0.3$~mb and $0.7$~mb tunings, so that the two configurations only differ once the hadronic phase begins. The formation-time delay $\Delta\tau(b)$ likewise leaves the parton-level dynamics untouched and enters only through the hadron formation time in the coalescence step. The fact that the residual $\sigma$-dependence of the final-state $\gamma$, together with the bulk of the CME dilution, is concentrated in the hadronic stage therefore identifies the ART hadron cascade in AMPT---rather than the initial condition or the parton cascade---as the dominant source of model dependence in our result.

A more definitive small-system CME prediction will accordingly require transport frameworks that go beyond the two-body hadronic scattering treatment of ART. Two directions look particularly promising: (i) improved partonic Boltzmann transport with running-coupling and $2\!\leftrightarrow\!3$ processes, along the lines of the BAMPS approach of Xu and Greiner~\cite{Xu:2004mz}, extended to include a chiral-anomaly source of the type used in the present work; and (ii) hybrid frameworks such as the CoLBT-hydro model~\cite{Chen:2017zte}, which couples a $3\!+\!1$D hydrodynamic evolution of the bulk QGP to a linear Boltzmann propagation of hard partons and a Cooper--Frye + hadron-cascade back end. Applying the same event-by-event CME source used here to such frameworks---while keeping the initial geometry and the $|\bB|$ distribution fixed---would isolate whether the strong hadronic-phase dilution seen in AMPT is model-generic or is dominated by ART-specific dynamics, and would sharpen the theoretical benchmark against which future PHENIX and sPHENIX $p^{\uparrow}+$Au CME searches can be interpreted.

\begin{acknowledgments}
We were supported by the NSFC under Projects No. 12075094. The computation is completed in the HPC Platform of Huazhong University of Science and Technology.
\end{acknowledgments}

\bibliographystyle{apsrev4-2}
\bibliography{references}

\end{document}